\documentclass[10pt,prb,superscriptaddress,amsmath,amssymb, twocolumn]{revtex4-1}

\usepackage{graphicx}
\usepackage{dcolumn}
\usepackage{bm}

\renewcommand{\vec}{\mathbf}

\begin{document}


\title{Effective potential and quantum criticality for imbalanced Fermi mixtures}

\author{Piotr Zdybel}
\author{Pawel  Jakubczyk}
\affiliation{
 Institute of Theoretical Physics, Faculty of Physics, University of Warsaw\\
 Pasteura 5, 02-093 Warsaw, Poland
}%

\date{June 12, 2018}

\begin{abstract}
We study the analytical structure of the effective action for spin- and mass-imbalanced Fermi mixtures at the onset of the superfluid state. Of our particular focus is the possibility of suppressing the tricritical temperature to zero, so that the transition remains continuous down to $T=0$ and the phase diagram hosts a quantum critical point. At mean-field level we analytically identify such a possibility in a regime of parameters in dimensionality $d=3$. In contrast, in $d=2$ we demonstrate that the occurrence of a quantum critical point is (at the mean-field level) excluded. We show that the Landau expansion of the effective potential remains well-defined in the limit $T\to 0^+$ except for a subset of model parameters which includes the standard BCS limit. We calculate the mean-field asymptotic shape of the  transition line. Employing the functional renormalization group framework we go beyond the mean field theory and demonstrate the stability of the quantum critical point in $d=3$ with respect to  fluctuations. 

\end{abstract}

\maketitle


\section{\label{sec:level1}Introduction} 
Mixtures of cold Fermi gases received enormous interest over the last years both from the experimental \cite{zwierlein_fermionic_2006, partridge_pairing_2006, bloch_many-body_2008,Ketterle_2009, ong_spin-imbalanced_2015, murthy_observation_2015, mitra_phase_2016}  and theoretical \cite{giorgini_theory_2008, gubbels_imbalanced_2013, chevy_ultra-cold_2010, bulgac_induced_2006, combescot_introduction_2007, radzi_07, radzihovsky_imbalanced_2010,Torma_2016,Turlapov_2017, Strinati_review_2018} points of view. This is on one hand triggered by the developments in controlled cooling of trapped atomic gases, and, on the other, 
by the theoretically predicted possibilities of realizing unconventional superfluid phases in such systems. The latter include, for example, the interior-gap (Sarma-Liu-Wilczek) superfluids \cite{sarma_influence_1963, liu_interior_2003} or the nonuniform Fulde-Ferrell-Larkin-Ovchinnikov (FFLO) states.\cite{fulde_superconductivity_1964, larkin_nonuniform_1965, Kinnunen_2018} The physics 
explored in this context is not specific to cold atomic gases, but finds close analogies in fields as distinct as the traditional solid-state physics,\cite{georges_condensed_2007, ho_quantum_2009, casalbuoni_inhomogeneous_2004} nuclear physics \cite{casalbuoni_inhomogeneous_2004,bailin_superfluidity_1984} or astrophysics of neutron star cores.\cite{bailin_superfluidity_1984,chamel_superfluidity_2017}  

An interesting question concerns the character of the superfluid transition at $T\to 0^+$. Such a transition can be tuned, for example, by manipulating the concentration of the different atomic species. As was recognized in a number of mean-field (MF)\cite{gubbels_sarma_2006, parish_finite-temperature_2007, Feiguin_2009, baarsma_population_2010, Caldas-2012, Klimin_2012, Kujawa_2011, Rosher_2015, Toniolo_17}  studies, it is rather generically of first order and becomes continuous only above a tricritical temperature $T_{tri}$ (see Fig.~1 for illustration). However, Ref.~\onlinecite{parish_polarized_2007} identified also a possibility of realizing a quantum critical point (QCP) as well as a quantum tricritical point at the mean-field level. The question concerning the actual 
order of the quantum phase transition is interesting since the occurrence of a quantum critical point and the related enhanced fluctuation effects feedback to the fermionic degrees of freedom (see e.g. Refs.~\onlinecite{Belitz_2005, Lohneysen_2007}). This leads to self-energy effects which may, for example, result in a breakdown of the quasiparticle concept and the occurrence of anomalous regions of the phase diagram both within the normal and the superfluid phases. The emergent physics has not as yet been fully explored. 

It is therefore interesting to understand the conditions under which the system in question may host a QCP. Most of the important earlier studies relied on numerical extraction of the MF free energy profiles leading to the phase diagrams. The present work contributes an analytical understanding of the structure of the effective action in the limit of low temperatures and gives criteria for the occurrence of the QCP for the spin and mass imbalanced systems. We precisely characterize the parameter region leading to the appearance of a QCP in $d=3$ at MF level and a phase diagram as illustrated in Fig.~2. We demonstrate that a QCP is (at MF level) excluded in $d=2$. Our study indicates that the Landau expansion of the effective potential remains well-defined down to 
$T=0$ except for a set of model parameters including the balanced (BCS) case, where the loop integrals defining the Landau coefficients diverge upon taking the limit $T\to 0^+$. Using a Sommerfeld-type expansion the shape of the transition line can be also calculated at $T>0$. For the mass-balanced case in $d=3$ the MF phase diagram was systematically analyzed in Ref.~\onlinecite{radzi_07}. In particular it was shown that a QCP may be realized on the BEC side. The mass-balanced case in $d=2$ was adressed analytically in Ref.~\onlinecite{Sheehy_15} pointing at a generically first-order transition between the normal and superfluid phases. In addition, that work discussed the Landau expansion within the FFLO phase finding nonanalytical contributions.

In several condensed-matter contexts,\cite{Belitz_2005, Lohneysen_2007} for example the quantum phase transitions in ferromagnets\cite{Belitz_02} or superconductors,\cite{Halperin_74, Li_09, Boettcher_18} one encounters the situation where the quantum phase transition is driven first-order by fluctuation effects. Using the functional renormalization-group framework, we investigate such a possibility in the presently considered context. The analysis performed in $d=3$ points at the robustness of the quantum critical point with respect to fluctuations. No indication of an instability towards a first-order transition is observed. 

The structure of the manuscript is as follows: in Sec.~II we introduce the considered model and its mean-field treatment leading to the expression for the free energy. In Sec.~III we analyze the Landau expansion and discuss its regularity in the limit $T\to 0^+$. The Landau coefficients are explicitly evaluated and analyzed in detail in the limit $T\to 0^+$ in Sec.~IV. In Sec.~V we employ the Sommerfeld expansion to address the asymptotic shape of the $T_c$-line. In Sec.~VI we discuss the effects expected beyond MF theory. In particular, we perform a functional renormalization-group calculation demonstrating the stability of the QCP obtained at the MF level with respect to fluctuations. In Sec.~VII we summarize the paper.

\section{\label{sec:level2}Model and mean-field theory}
We consider a two-component fermionic mixture characterized by distinct particle masses and concentrations which may act as tuning-parameters. The inter-species attractive contact interaction triggers $s$-wave pairing. The Hamiltonian reads   
\begin{eqnarray} 
\label{Hamiltonian}
\hat{H}-\sum_{\sigma}\mu_\sigma \hat{n}_\sigma=\sum_{\vec{k},\sigma}\xi^{\phantom{\dagger}}_{\vec{k},\sigma}c^\dagger_{\vec{k},\sigma}c^{\phantom{\dagger}}_{\vec{k},\sigma}+\nonumber\\ 
+\frac{g}{V}\sum_{\vec{k},\vec{k}',\vec{q}}c^\dagger_{\vec{k}+\vec{q}/2,\uparrow}c^\dagger_{-\vec{k}+\vec{q}/2,\downarrow}c^{\phantom{\dagger}}_{\vec{k}'+\vec{q}/2,\downarrow}c^{\phantom{\dagger}}_{-\vec{k}'+\vec{q}/2,\uparrow}\;,
\label{1}
\end{eqnarray}
where $\sigma\in\{\uparrow, \downarrow\}$ labels the particle species, $\xi_{\vec{k},\sigma}=\vec{k}^2/2m_\sigma -\mu_\sigma$ are the dispersion relations, $g<0$ is the interaction coupling and $V$ denotes the volume of the system. In general, the masses $m_\sigma$ and chemical potentials $\mu_\sigma$ corresponding to the distinct species can be different. Shifting the imbalance parameter $h=(\mu_\uparrow-\mu_\downarrow)/2$ away from zero mismatches the Fermi surfaces and suppresses superfluidity. The quantity $h$ therefore constitutes a natural non-thermal control parameter to tune the system across the superfluid quantum phase transition. 
\begin{figure}
\centering 
    \includegraphics[width=0.45 \textwidth]{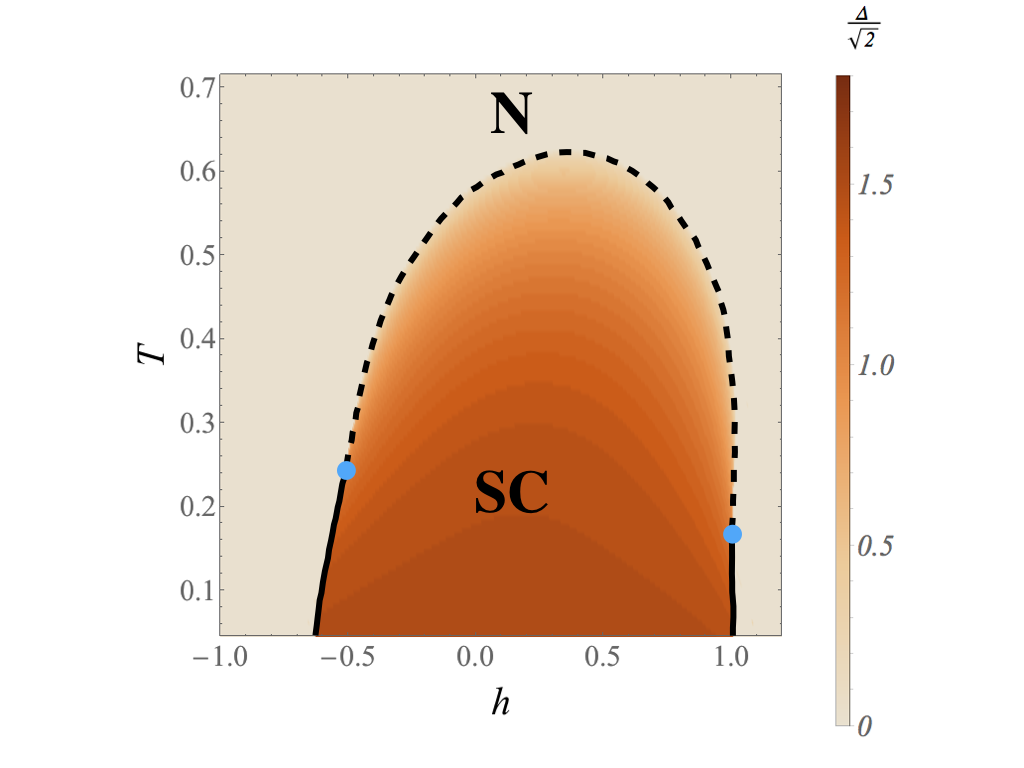} 
    \caption{ A typical mean-field phase diagram in $d=3$. The low-temperature superfluid phase is separated from the normal phase with a first-order phase transition (bold line) at $T$ sufficiently low, and with a second-order phase transition (dashed line) at $T$ higher. The blue dots indicate the tricritical points. The colors refer to the value of the order parameter $(\Delta)$. The plot parameters are $r=m_\downarrow/m_\uparrow=2$, $\mu=0.1$, $g=-1.7$, and $\Lambda=10$.}
    \label{pd}
\end{figure} 
\begin{figure}
\centering 
    \includegraphics[width=0.45 \textwidth]{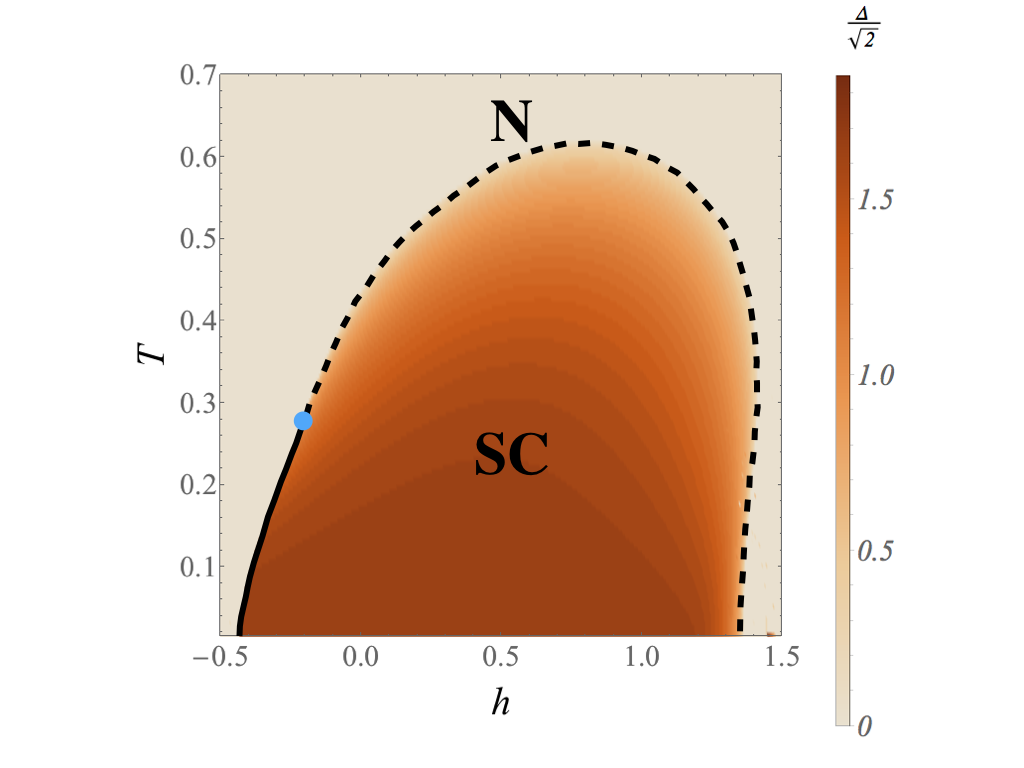} 
    \caption{ A mean-field phase diagram in $d=3$ displaying a quantum critical point. The colors refer to the value of the order parameter $(\Delta)$. The plot parameters are $r=m_\downarrow/m_\uparrow=5$, $\mu=0.1$, $g=-1.4$, and $\Lambda=10$.}
    \label{pdqcp}
\end{figure}
The mean-field phase diagram of the system defined by Eq.~(\ref{Hamiltonian}) was addressed in a sequence of studies spread over the last years. In addition to the normal and uniform superfluid phases (as shown in Fig.~1) it displays a tiny region hosting the FFLO state. This phase is fragile to fluctuation effects and it remains open under what conditions such a superfluid may be realized. In the present study the FFLO state will be disregarded. For discussions concerning its stability in the context of cold atoms see Refs.~\onlinecite{Shimahara_1998, Samokhin_2010, Radzihovsky_2011, Yin_2014, Jakubczyk_2017, Ptok_17} .  

Assuming $s$-wave pairing at ordering wavevector ${\bf q}=0$, the mean-field grand-canonical potential $\omega(T, \mu, h)$ may be derived along the standard track. It reads:
\begin{eqnarray}
\omega(T, \mu, h)=\min\left[\omega_L(\Delta)\right] = \min_\Delta\Bigg\{-\frac{|\Delta|^2}{g}+\nonumber\\-\frac{1}{\beta}\int_\vec{k}\sum_{i\in\{+,-\}}\ln\left(1+\mathrm{e}^{-\beta E^{(i)}_\vec{k}}\right)\Bigg\}, 
\label{2}
\end{eqnarray} 
where $\Delta$ is the superfluid order parameter field and $\int_\vec{k}(\cdot)=\int\frac{\mathrm{d}^d \vec{k}}{(2\pi)^d}(\cdot)$. The elementary excitations' energies are given by:
\begin{eqnarray}
E^{(\pm)}_\vec{k}=\frac{\xi_{\vec{k},\uparrow}-\xi_{\vec{k},\downarrow}}{2}\pm\sqrt{\xi_\vec{k}^2+|\Delta|^2}, 
\label{3}
\end{eqnarray}
where $\xi_\vec{k}=(\xi_{\vec{k},\uparrow}+\xi_{\vec{k},\downarrow})/2$. We use $\mu=(\mu_\uparrow+\mu_\downarrow)/2$ as the average chemical potential, while the average "Zeeman" field, which describes spin imbalance is denoted as $h=(\mu_\uparrow-\mu_\downarrow)/2$. By minimizing Eq.~(\ref{2}) one determines the grand-canonical potential together with the superfluid order parameter expectation value. We show illustrative profiles of $\omega_L(\Delta)$ in Fig.~3. The normal phase (N) corresponds to $\Delta=0$ and is separated from the superfluid (SF) phase characterized by $|\Delta|>0$ with a phase transition. The latter is typically of first order for $T$ sufficiently low and becomes continuous above the tricritical temperature $T_{tri}$.  Numerical minimization of Eq.~(\ref{2}) (or equivalent expressions) constituted the basis of the earlier studies. Such analysis  typically lead to  phase diagrams as exemplified in Fig.~1. We show however that Eq.~(2) is also susceptible to an analytical treatment which gives additional insights and allows for making some exact and general statements at the mean-field level.
\begin{figure}
\centering
    \includegraphics[width=0.45 \textwidth]{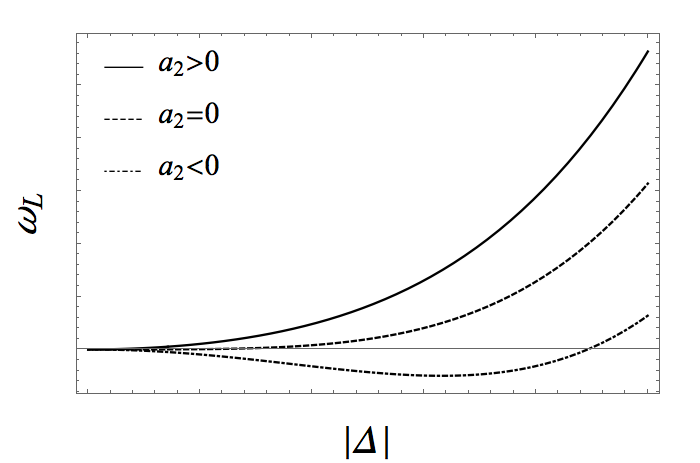}
    \caption{Schematic illustration of the effective potential $\omega_L(\Delta)$  for positive $a_4$. In such a case $a_2>0$ correspond to normal phase, $a_2<0$ to the superfluid phase and $a_2=0$  to the 2nd order phase transition. }
    \label{lf}
\end{figure}
\section{\label{sec:level2}Landau expansion}
The Landau theory of phase transitions postulates an analytical expansion of $\omega_L(\Delta)$:  
\begin{eqnarray}
\omega_L(\Delta)=\omega_0+a_2|\Delta|^2+a_4|\Delta|^4+a_6 |\Delta|^6 +\dots, 
\label{4}
\end{eqnarray}
where we take only even powers of the order parameter, preserving the $U(1)$ symmetry. 
The Landau coefficients $a_i$ are functions of the system parameters and the thermodynamic fields. For the present case they may be extracted by taking consecutive derivatives of $\omega_L(\Delta)$  given by Eq.~(\ref{2}) and evaluating at $\Delta=0$.
The coefficient $a_2$ follows from:
\begin{eqnarray}
a_2=\left(\frac{\partial \omega_L}{\partial|\Delta|^2}\right)_{|\Delta|^2=0}=\nonumber\\=-\frac{1}{g}-\frac{1}{4}\int_\vec{k}\frac{1}{\xi_{\vec{k}}}\sum_\sigma\tanh\left(\frac{\beta\xi_{\vec{k},\sigma}}{2}\right)\;,
\label{5}
\end{eqnarray}
while for the quartic coefficient we obtain
\begin{widetext}
\begin{eqnarray}
a_4=\frac{1}{2}\left(\frac{\partial^2{\omega}_L}{\partial\left(|\Delta|^2\right)^2}\right)_{|\Delta|^2=0}=\frac{1}{16}\int_\vec{k}\frac{1}{\xi_{\vec{k}}^3}\sum_\sigma\left[\tanh\left(\frac{\beta\xi_{\vec{k},\sigma}}{2}\right)-\frac{\beta\xi_{\vec{k}}}{2}\mathrm{cosh}^{-2}\left(\frac{\beta\xi_{\vec{k},\sigma}}{2}\right)\right].
\label{6}
\end{eqnarray}
\end{widetext} 
The higher-order coefficients may be derived by differentiating Eq.~(\ref{2}) further. 
The Landau coefficient $a_i$ may also be understood as a Fermi loop with $i$ external (bosonic) legs evaluated at the (external) momenta zero. The Fermi propagators are gapped by the lowest Matsubara frequency which vanishes for $T\to 0^+$. It is therefore not immediately obvious, under which circumstances the loop integrals converge for $T\to 0^+$ [i.e. the expressions given by Eq.~(\ref{5}) and Eq.~(\ref{6}) remain finite for $T\to 0^+$]. Potential problems of this nature occur at any dimensionality and are rather clearly visible in Eq.~(\ref{5}) and Eq.~(\ref{6}). For example, by specifying to the balanced case $\xi_{\vec{k},\uparrow}=\xi_{\vec{k},\downarrow}=\xi_{\vec{k}}$ we easily realize that the coefficient $a_2$ [as given by Eq.~(\ref{5})] contains a contribution  which diverges for $\beta\to\infty$. This signals the breakdown of the Landau expansion of Eq.~(\ref{4}) for $T\to 0^+$ in the balanced case. On the other hand, one may evaluate the $T\to 0^+$ limit of Eq.~(\ref{5}) (see Sec.~IV) and obtain generically finite expressions for the imbalanced case. 

The analysis of the quartic coupling [Eq.~(\ref{6})] is slightly more complex. Since potential divergencies in Eq.~(\ref{6}) come from the vicinity of $\xi_{\vec{k}}=0$, we restrict the integration region in Eq.~(\ref{6}) to a shell  of width $2\epsilon$ around $\xi_{\vec{k}}=0$. Upon expanding the integrands, performing the integrations, and, at the end, considering $T\to 0^+$, we find that the limit is finite provided 
\begin{equation} 
\label{Landau_reg}
\mu \neq h \frac{r+1}{r-1}\;,
\end{equation}
where we introduced $r=\frac{m_{\downarrow}}{{m_\uparrow}}$ and assumed $m_\downarrow>m_\uparrow$.  The analysis can be extended to higher Landau coefficients. As a result we obtain that Eq.~(\ref{Landau_reg}) gives a (necessary and sufficient) condition for the regularity of the Landau expansion Eq.~(\ref{4}) in the limit $T\to 0^+$. The above result does not depend on the system dimensionality. For fixed $r$ Eq.~(\ref{Landau_reg}) describes a straight line in the $(h,\mu)$ plane, whose slope diverges for equal particle masses ($r\to 1^+$).  We also observe that the standard balanced case corresponds to the limit $h\to 0$ and $r\to 1^+$, which, from the point of view of Eq.~(\ref{Landau_reg}) is not defined.  In this case, the way of taking the limits selects the point on the half-line ($\mu>0$, $h=0$). The condition in Eq. (\ref{Landau_reg}) corresponds to a situation where the two Fermi surfaces coincide.

Here we also point out that, provided the expansion of Eq.~(\ref{4}) exists, the condition for a continuous transition reads 
\begin{equation}
a_2 =0,\;\; a_4>0\;,   
\label{CP_condition}
\end {equation}
while a tricritical point occurs iff 
\begin{equation}
a_2 =0,\;\; a_4=0,\;\; a_6>0\;.   
\label{TP_condition}
\end {equation}

\section{\label{sec:level2}Zero temperature}
\begin{figure}
\centering
    \includegraphics[width=0.5 \textwidth]{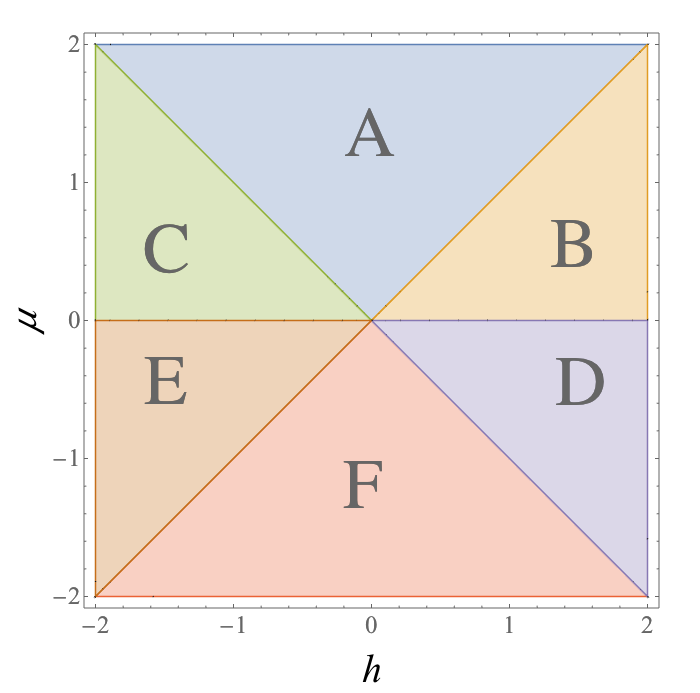}
    \caption{Distinct regions of the phase diagram at $T=0$.  (A) $\mu_\uparrow>0$, $\mu_\downarrow>0$; (B) $\mu>0$, $\mu_\uparrow>0$, $\mu_\downarrow<0$; (C) $\mu>0$, $\mu_\uparrow<0$, $\mu_\downarrow>0$; (D) $\mu<0$, $\mu_\uparrow>0$, $\mu_\downarrow<0$; (E) $\mu<0$, $\mu_\uparrow<0$, $\mu_\downarrow>0$ and (F) $\mu_\uparrow<0$, $\mu_\downarrow<0$.}
    \label{ar}
\end{figure}

We now consider the limiting form of the expressions given by Eq.~(\ref{5}) and Eq.~(\ref{6}) for $T\to 0^+$. Using  $\tanh\left(\frac{\beta\xi_{\vec{k},\sigma}}{2}\right)=1-2f(\xi_{\vec{k},\sigma})$, where $f(\cdot)$ is the Fermi function, we find
\begin{eqnarray}
a_2^{(0)} = \lim_{T\to 0^+} a_2=-\frac{1}{g}-\frac{1}{2}\int_\vec{k}\frac{1}{\xi_\vec{k}}\left[1-\sum_\sigma\theta\left(-\xi_{\vec{k},\sigma}\right)\right],
\label{7}
\end{eqnarray}
where $\theta(\cdot)$ is the Heaviside step function. Similarly, taking advantage of the fact that $\frac{\mathrm{d}}{\mathrm{d}\xi_{\vec{k},\sigma}} \tanh\left(\frac{\beta\xi_{\vec{k},\sigma}}{2}\right)=\frac{\beta}{2}\mathrm{cosh}^{-2}\left(\frac{\beta\xi_{\vec{k},\sigma}}{2}\right)$ we
 obtain the corresponding expression for the quartic coupling 
 \begin{eqnarray}
a_4^{(0)}=\frac{1}{8}\int_\vec{k}\frac{1}{\xi_\vec{k}^3}\left[1-\sum_\sigma\left\{\theta\left(-\xi_{\vec{k},\sigma}\right)+\xi_\vec{k}\delta\left(\xi_{\vec{k},\sigma}\right)\right\}\right],
\label{8}
\end{eqnarray}
where $\delta(\cdot)$ is the Dirac delta. The integrations may be performed analytically, however, their form depends on the dimensionality $d$. We discuss the two physically most relevant cases of $d=2$ and $d=3$ separately. The analysis requires dividing  the $(h,\mu)$ plane into several subsets as illustrated in Fig.~4. This complication arises because zeros of the Heaviside and Dirac distributions in Eq.~(\ref{7}) and Eq.~(\ref{8}) may lie either inside or outside the integration domains. In consequence, we are led to considering six distinct regions shown in the Fig. \ref{ar}. Their physical significance is discussed in detail in  Sec.~IVC. We note however already at this point that physically interesting parameter regions occur also for negative values of $\mu_\sigma$ - see Sec.~IVC. 

\subsection{d=2} 
We now analyze the expressions given by Eq.~(\ref{7}) and Eq.~(\ref{8}) for $d=2$.
\subsubsection{Coefficient $a_2^{(0)}$}
As already explained, the analysis requires considering the distinct regions described in Fig.~4 separately. By doing the integral in Eq.~(\ref{7}) in regime A, we obtain the following expression:
\begin{eqnarray}
a^{(A)}_2=-\frac{1}{g}-\frac{m_r}{4\pi}\ln\left[\frac{|\Lambda^2-2\mu m_r|\cdot2\mu m_r}{\prod_{\sigma}|\lambda_{\sigma}^2-2\mu m_r|}\right],
\label{12}
\end{eqnarray}
where $\lambda_{\sigma}^2=2\mu_\sigma m_\sigma$, $m_r=\frac{2m_\uparrow m_\downarrow}{m_\uparrow+m_\downarrow}$ and $\Lambda$ denotes the upper momentum cutoff. Similarly, for the regions B, C, D, and E, we have:
\begin{eqnarray}
a^{(B-E)}_2=-\frac{1}{g}-\frac{m_r}{4\pi}\ln\left[\frac{|\Lambda^2-2\mu m_r|}{|\lambda_{\sigma}^2-2\mu m_r|}\right]\;, 
\label{13}
\end{eqnarray} 
with $\sigma=\uparrow$ in regimes B, D, and $\sigma=\downarrow$ in regimes C and E. 

Finally, for the region F the Landau coefficient $a_2^{(0)}$ is given by:
\begin{eqnarray}
a^{(F)
}_2=-\frac{1}{g}-\frac{m_r}{4\pi}\ln\left[\frac{|\Lambda^2-2\mu m_r|}{|2\mu m_r|}\right].
\label{14}
\end{eqnarray}
The coefficient $a_2^{(0)}$ must vanish at the QCP according to Eq.~(\ref{CP_condition}). In Eqs. (\ref{12}-\ref{14}), the contribution involving the logarithm is negative (provided $\Lambda$ is sufficiently large). The attractive interaction coupling $g<0$ can therefore be tuned so that $a_2^{(0)}$ is zero. In an experimental situation this is achievable via Feshbach resonances.\cite{Feshbach_paper} Nevertheless, as we show below, the coefficient $a_4^{(0)}$ is generically negative in $d=2$ which renders the transition necessarily first order. 
\subsubsection{Coefficient $a_4^{(0)}$}
In analogy to the above analysis of the  coefficient $a_2^{(0)}$, we consider the different parameter space regions illustrated in Fig.~4 and evaluate the Landau coefficient $a_4^{(0)}$ from Eq.~(\ref{8}) in $d=2$. Within the region A ($\mu_\uparrow>0$, $\mu_\downarrow>0$) we obtain:
\begin{eqnarray}
a_4^{(A)}=-\frac{m_r^3}{8\pi}\Bigg[\frac{1}{\left(\Lambda^2-2\mu m_r\right)^2}+\frac{1}{\left(2\mu m_r\right)^2}+\\+\sum_{\sigma}\frac{(m_\sigma/m_{\bar{\sigma}})}{\left(\lambda_\sigma^2-2\mu m_r\right)^2}\Bigg]. \nonumber
\label{15}
\end{eqnarray} 
We introduced the index $\bar{\sigma}$ denoting the species opposite to $\sigma$ [i.e for $\sigma=\uparrow$ we have $\bar{\sigma}=\downarrow$ and vice versa]. 
For the subsets B, C, D, and E we find:
\begin{eqnarray}
a_4^{(B-E)}=-\frac{m_r^3}{8\pi}\left[\frac{1}{\left(\Lambda^2-2\mu m_r\right)^2}+\frac{(m_\sigma/m_{\bar{\sigma}})}{\left(\lambda_\sigma^2-2\mu m_r\right)^2}\right]\;,
\label{16}
\end{eqnarray} 
with $\sigma=\uparrow$ in regimes B, D, and $\sigma=\downarrow$ in regimes C and E. 

Finally, for the region F ($\mu_\uparrow<0$, $\mu_\downarrow<0$) we have:
\begin{eqnarray}
a_4^{(F)}=-\frac{m_r^3}{8\pi}\left[\frac{1}{\left(\Lambda^2-2\mu m_r\right)^2}-\frac{1}{\left(2\mu m_r\right)^2}\right].
\end{eqnarray}
With the exception of $a_4^{(F)}$ the above expressions are manifestly negative. Within regime (F) we observe that the expressions for $a_2^{(F)}$ and $a_4^{(F)}$ involve no dependence on $h$ and therefore $\Delta$ remains constant if $h$ is varied (at constant $\mu$). This implies that no phase transition (first or second order) is possible within the parameter regime (F).  Therefore, Eq.~(\ref{CP_condition}) is never fulfilled.

We conclude that the occurrence of a QCP is generally ruled out for $d=2$ at the MF level. Similar results for the mass-balanced case were obtained by Sheely\cite{Sheehy_15}, where it was pointed out that the phase transition at T=0 between the normal and superfluid phases is first-order.

\subsection{d=3}
The study in $d=3$ parallels the above analysis in $d=2$. The parameter space is again split into the distinct regions depicted in Fig.~4.  Evaluating the integrals in Eq.~(\ref{7}) and Eq.~(\ref{8}) yields the Landau coefficients $a_2^{(0)}$ and $a_4^{(0)}$. Subsequently we check if the condition for the occurrence of the QCP [Eq.~(\ref{CP_condition})] can be fulfilled. We relegate the obtained expressions for $a_2^{(0)}$ and $a_4^{(0)}$ to the Appendix and discuss the conclusions below.
\subsubsection{Coefficient $a_2^{(0)}$}
Similarly, as for the $d=2$ case, the coefficient $a^{(0)}_2$ contains a positive contribution related to the coupling constant $g$ and a negative part dominated by a term proportional to $\Lambda$ [Eqs.~(\ref{A1})-(\ref{A4})]. As a result, the interaction strength can be tuned such that $a_2^{(0)}$ equal zero. The coefficient $a_4^{(0)}$ carries no dependence on $g$.
\subsubsection{Coefficient $a_4^{(0)}$}
The expressions for $a_4^{(0)}$ are given in Eqs.~(\ref{B1})-(\ref{B4}). The analysis shows that the sign of $a_4^{(0)}$ is negative in regime (A) and  positive in regime (F). Alike for $d=2$ the dependence of both $a_2^{(0)}$ and $a_4^{(0)}$ on $h$ drops out within regime (F),  which excludes a phase transition for $h$ falling therein. For small mass imbalance ($r\approx 1$) the region corresponding to positive $a_4^{(0)}$ occupies the set (F) and tiny regions of regimes (D) and (E). Upon increasing $r$, the $a_4^{(0)}>0$ region covers increasingly large portions of the region (D), and, for $r>3.01$ it intrudes into regime (B). In the limit $r\to\infty$, the region with $a_4^{(0)}>0$ fully covers the regimes (F), (D), and (B). The evolution of the subset of the $\mu -h$ plane characterized by $a_4^{(0)}>0$ upon varying $r$ is depicted in Fig.~ 5. 
\begin{figure}
\centering 
    \includegraphics[width=0.3 \textwidth]{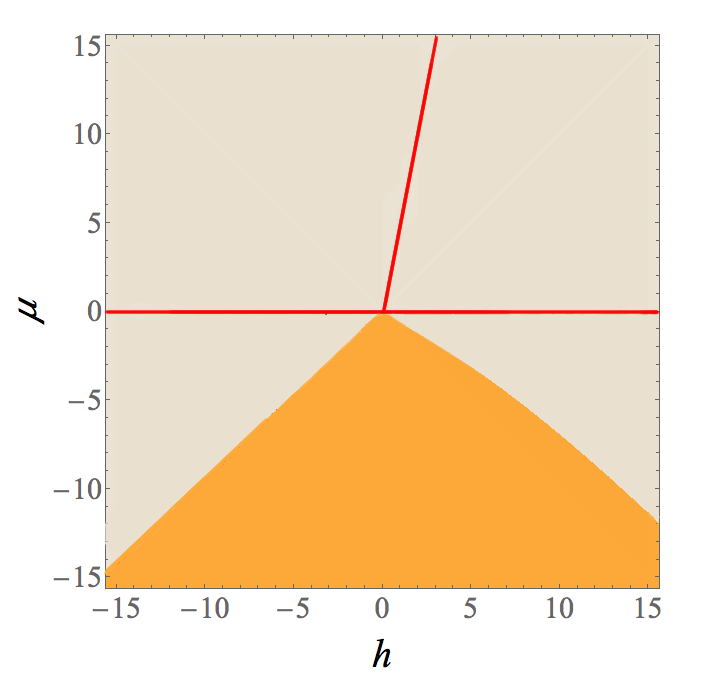}
    \includegraphics[width=0.27 \textwidth]{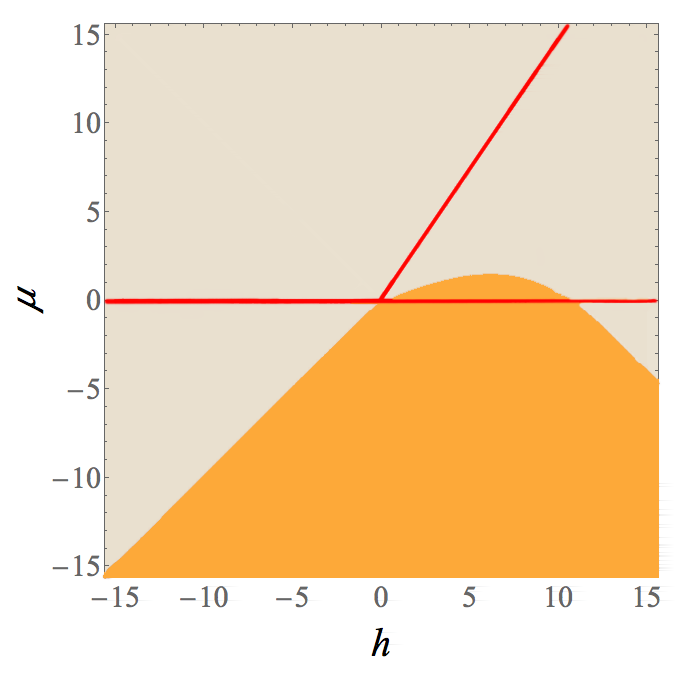}
    \includegraphics[width=0.3 \textwidth]{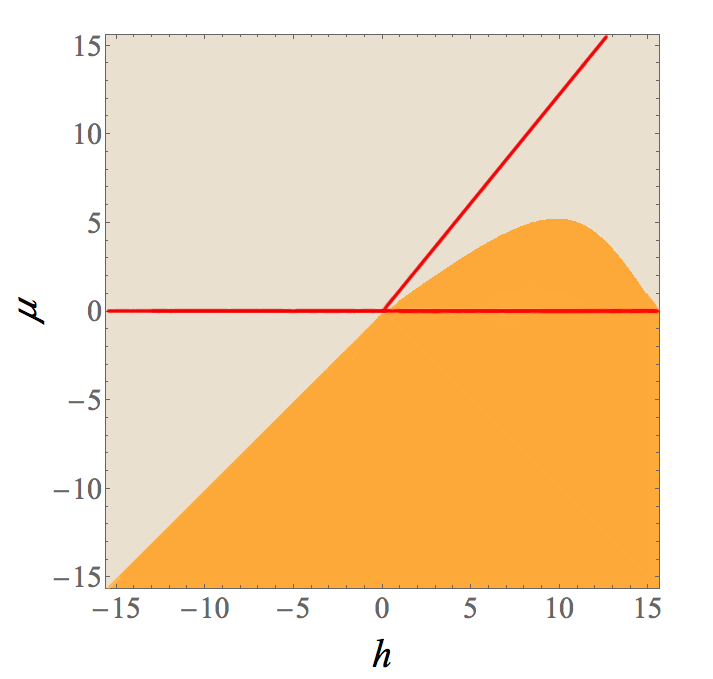}
    \caption{Evolution of the subset of the $\mu -h$ plane characterized by $a_4^{(0)}>0$ upon varying $r$. The (light) beige area corresponds to negative $a_4^{(0)}$, while in the (darker) orange area $a_4^{(0)}>0$. The coefficient $a_4^{(0)}$ is singular along the red straight lines. The first  diagram corresponds to $r=1.5$, the second one to $r=5$, the last one to $r=10$. Upon increasing $r$ towards $r\to \infty$ the orange region extends further to cover half of the $\mu -h$ plane located below the diagonal.   }
    \label{a4}
\end{figure}
The emergent picture is very different as compared to that obtained for $d=2$, where we showed that a QCP is completely excluded (at MF level). In $d=3$ the possibility of realizing a second-order transition at $T=0$ turns out to be restricted to situations, where one of the chemical potentials $\mu_\sigma$ is negative. In the next section we reinterpret the problem using the densities $n_\sigma$ instead of $\mu_\sigma$ as the control parameters. We show that (due to interaction effects) positive densities may well correspond to negative chemical potentials (also at $T=0$).  

Above we restricted to $r\geq 1$. For $r\in ]0,1[$ the emergent picture is analogous but $h\leftrightarrow -h$. 

\subsection{Particle densities}
In this section we discuss the physical significance of the regions considered in Fig. 4. This requires resolving the relation between the particle densities $n_\sigma$ and the chemical potentials $\mu_\sigma$.

We begin by considering a reference situation where $\Delta=0$ for all possible values of $\mu$ and $h$, which corresponds to the non-interacting two-component Fermi mixture ($g=0$). The relation between $n_\sigma$ and $\mu_\sigma$ is then given (at $T=0$)  by 
\begin{eqnarray}
n_\sigma=\frac{(m_\sigma\mu_\sigma)^{d/2}\theta(\mu_\sigma)}{(2\pi)^{d/2}\Gamma\left(\frac{d}{2}+1\right)}\;.
\label{19}
\end{eqnarray} 
A graph showing this dependence is presented in Fig.~6 for $\mu>0$ and $\mu<0$. Obviously, the species $\sigma$ is expelled from the system if $\mu_\sigma<0$. 
\begin{figure}[h]
\centering
    \includegraphics[width=0.45 \textwidth]{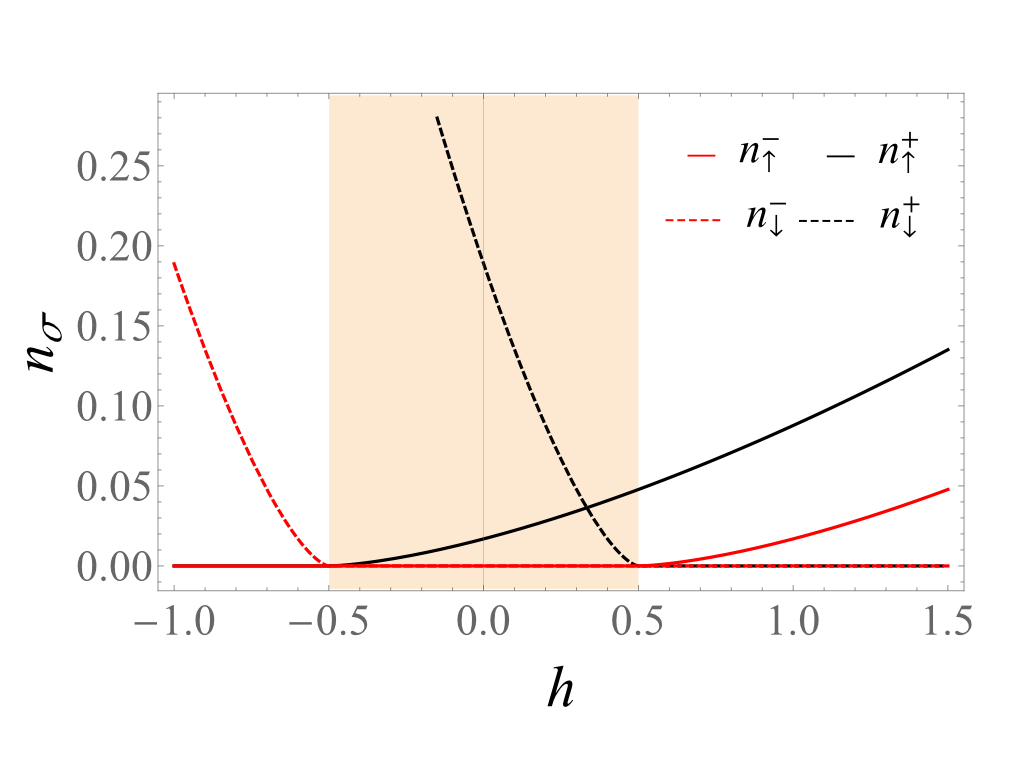}
    \caption{Particle densities $n_\sigma$ for the non-interacting two-component Fermi mixture ($g=0$) as a function of $h$ for $d=3$, where we fixed $T=0$ and $r=5$. Here the black curves $n_\sigma^+$ correspond to $\mu=0.5$, while the red lines $n_\sigma^-$ correspond to $\mu=-0.5$. In the shaded area both species are present for $\mu=0.5$, but none at $\mu=-0.5$.}
    \label{obs1}
\end{figure}

Now consider $g<0$. Identifying $\sigma=\uparrow$ with $+1$ and $\sigma =\downarrow$ with $-1$, the density of the species $\sigma$ is given by:
\begin{eqnarray}
n_\sigma(\Delta)=\frac{1}{V}\sum_{\vec{k}}\Big[|u_{\vec{k}}|^2f\left(\sigma E_{\vec{k}}^{(\sigma)}\right)+\nonumber\\+|v_{\vec{k}}|^2\left(1-f\left(\bar{\sigma}E_{\vec{k}}^{(\bar{\sigma})}\right)\right)\Big],
\label{18}
\end{eqnarray}
where $\bar{\sigma}=-\sigma$, $|u_{\vec{k}}|^2+|v_{\vec{k}}|^2=1$ and $|u_{\vec{k}}|^2=\frac{1}{2}\left[1+\left(\xi_{\vec{k}}/\sqrt{\xi^2_{\vec{k}}+|\Delta|^2}\right)\right]$. One may now fix $\mu$ and $h$, compute $\Delta$ (see Sec.~II) and use Eq.~(\ref{18}) to extract $n_\sigma$. We first observe that for $\Delta=0$ and $T=0$ Eq.~(\ref{18}) reduces to Eq.~(\ref{19}). This implies that any conceivable phase transition at $\mu_\uparrow<0$ or $\mu_\downarrow<0$  occurs between the superfluid and a fully polarized gas. In particular, in view of the results of Sec.~IVB, we conclude that all possible QCPs in $d=3$ fall into this category. Generically, $\Delta>0$ implies the presence of particles of both species in the system. In consequence, a phase transition at $\mu_\downarrow<0$ (or $\mu_\uparrow <0$) requires that the density of one of the species raises from zero to a finite value (either continuously or discontinuously, depending on the order of the transition). This is illustrated in Fig.~7, where we plot the densities $n_\uparrow$ and $n_\downarrow$ as function of $h$ for two values of $\mu$ (one positive and one negative). 
\begin{figure}[!]
\centering
    \includegraphics[width=0.45 \textwidth]{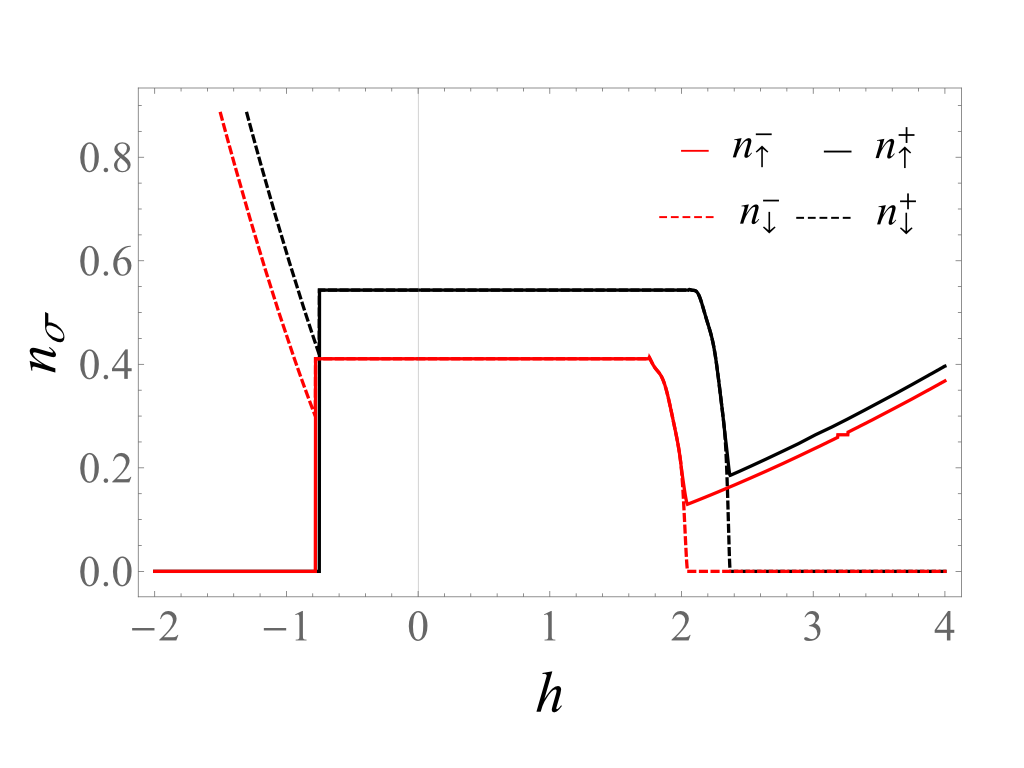}
    \caption{Particle densities $n_\sigma$ plotted as functions of $h$ for $d=3$, where we fixed $T=0$, $r=5$ and $g=-1.5$. The black curves $n_\sigma^+$ correspond to $\mu=0.1$ while the red lines $n_\sigma^-$ correspond to $\mu=-0.1$. For each of the values of $\mu$ one observes a first-order transition around $h\approx -1$, where $n_\uparrow$ increases from zero to a finite value, and a second-order transition around  $h\approx 2$, where $n_\downarrow$ continuously decreases to zero. In the normal phase the densities $n_\sigma$ follow the power law given by Eq.~(\ref{19}) and illustrated in Fig.~\ref{obs1}. The illustrated phase transitions are located regimes C and E ($h\approx -1$) and B and D ($h\approx 2$) (compare Fig.~4). Inspection of Fig.~5 (the plot in the middle) shows that the transitions at $h\approx -1$ correspond to $a_4^{(0)}<0$, while the transitions  at $h\approx 2$ are characterized by $a_4^{(0)}>0$.  }
    \label{obs2}
\end{figure}
The plot parameters are chosen so that (for each of the considered values of $\mu$) we encounter a first order phase transition at $h<0$ (corresponding to regimes C and E in Fig.~4) and a second-order transition at $h>0$ (corresponding to regimes B and D in Fig.~4). 

Summarizing the major conclusions of this section: we have shown that at mean-field level and $T=0$ the superfluid transition is inevitably first order in $d=2$. For $d=3$ we have demonstrated that a second-order quantum phase transition is possible only between a fully-polarized gas and the superfluid phase. Such a scenario is favorable at large mass imbalance ($r \gg 1$ or $r\ll 1$). Note however that the QCP exists even for $r=1$ on the BEC side of the BCS-BEC crossover (see  Ref.~\onlinecite{radzi_07}).
\section{\label{sec:level4}Finite temperature}
The numerical evaluation of the MF phase diagram (see Fig.~2) shows that the ordered phase extends when the temperature is increased from zero to finite values in the vicinity of the QCP (i.e. the slope of the $T_c$-line is positive for sufficiently low $T$). Here we analyze the asymptotic shape of the $T_c$ line in the vicinity of the QCP. The behavior observed in Fig.~2 can be understood employing the Sommerfeld (low-temperature) expansion\cite{kardar_statistical_2007} for the coefficient $a_2$ [Eq.  (\ref{5})]. We focus on regime B (see Fig. 4), which corresponds to the QCP depicted in Fig.~2. We fix $\mu$ and $r$ and perform the low-temperature expansion of Eq. (\ref{5}). We obtain: 
\begin{eqnarray}
a_2(T,h)=a_2^{(0)}(h)-\alpha(h) T^2+\dots,
\label{20}
\end{eqnarray}
where the coefficient $\alpha(h)$ is given by:
\begin{eqnarray}
\alpha(h)=\frac{ m_r m^2_\uparrow(\lambda^2_\uparrow+2\mu m_r)}{12\lambda_\uparrow(\lambda^2_\uparrow-2\mu m_r)^2}.
\label{21}
\end{eqnarray}
The first term in the Sommerfeld expansion corresponds to the zero-temperature Landau coefficient given by  Eq. (\ref{A2}) and the second term is the low-temperature correction. We expand $a_2^{(0)}$ around the ($T=0$) critical value $h_c$ of the field $h$ and find $h_c$  from the condition $a_2^{(0)}(h_c)=0$. This yields: 
\begin{equation}
T_c (h_c+\delta h)\approx\sqrt{\frac{\partial_h a_2^{(0)}|_{h=h_c}\delta h}{\alpha(h_c)}}\propto \sqrt{h-h_c},
\label{22}
\end{equation}
where $ \delta h $ is a small deviation from $ h_c $. The MF $T_c$-line is described by a power law with the exponent $1/2$, which is a generic value for Fermi systems. Notably $\delta h$ is positive, in agreement with the numerical results [for example Fig.~(2)]. 

\section{\label{sec:level4}Functional renormalization} 
The above analysis is restricted to the mean-field level. In the present section we employ the functional renormalization group (RG) framework to discuss fluctuation effects. As we already noted, in a number of condensed-matter contexts one encounters the effect of fluctuation driven first-order quantum phase transitions.\cite{Belitz_2005, Lohneysen_2007} Well-recognized examples include the ferromagnetic quantum phase transition\cite{Belitz_02} or the $s$-wave\cite{Halperin_74} as well as $p$-wave\cite{Li_09} superconductors. In the case of itinerant ferromagnets the transition is first-order at $T=0$ due to a term $\sim \phi^4\log\phi$ appearing in the effective action upon integrating out the (gapless) fermionic degrees of freedom. A different kind of nonanalyticity of the effective action is generated in the case of superconductors due to the coupling between the order parameter and the electromagnetic vector potential. We argue that no such mechanism is active for the presently discussed system defined in Sec.~II. By an explicit functional RG calculation (retaining terms up to infinite order in $\Delta$) we show that the QCP obtained at the MF level in the preceding sections for $d=3$ is stable with respect to the order-parameter fluctuations. We additionally note that the possibility of changing the order of quantum phase transitions from first to second due to order parameter fluctuations was demonstrated for effective bosonic field theories\cite{Jakubczyk_2009_phi6, Jakubczyk_2010} as well as specific microscopic fermionic models.\cite{Jakubczyk_2009_PRL, Yamase_2011_PRB, Yamase_2015, Boettcher_2015_PRA, Boettcher_2015_Phys_lett, Classen_2017}  Also (as is indicated by our analysis) in the present situation one anticipates the fluctuations to round the transition rather that drive it first order. Note that functional RG was previously employed to obtain the phase diagram in the mass balanced case (Ref.~\onlinecite{Boettcher_2015_Phys_lett}) and to study the imbalanced unitary Fermi mixtures (Ref.~\onlinecite{Rosher_2015}).

Our present analysis is restricted to $d=3$ and proceeds along the line analogous to Ref.~\onlinecite{Strack_2014}, where the possibility of driving the quantum phase transition second-order by fluctuations was discussed for $d=2$. We also observe, that a similar framework was employed in Ref.~\onlinecite{Boettcher_2015_PRA} for the presently discussed model in $d=3$ strictly at the unitary point with the conclusion that the transition is first order both at MF level and after accounting for fluctuations.  

Following Ref.~\onlinecite{Strack_2014} we integrate the order-parameter fluctuations by the flow equation for the effective potential: 
\begin{equation}
\label{LPA_flow}
\partial_\kappa U_\kappa(\rho) = \frac{1}{2}T\sum_{\omega_n}\int\frac{d^3 q}{(2\pi)^3}\partial_\kappa R_\kappa (\omega_n,\vec{q}) (G_L^R +G_T^R)\;,
\end{equation}
where $\omega_n$ are the (bosonic) Matsubara frequencies, $\rho=\frac{1}{2}\Delta^2$, while $G_L^R$ and $G_T^R$ denote the longitudinal and transverse $\rho$-dependent propagators: 
\begin{eqnarray}
G_L^R = \frac{1}{D} \left[Z_\omega \omega_n^2+Z\vec{q}^2+M_L^R\right]  \nonumber \\
G_T^R = \frac{1}{D} \left[Z_\omega \omega_n^2+Z\vec{q}^2+M_T^R\right]\;,
\end{eqnarray}
with $M_L^R=U'(\rho)+2\rho U''(\rho)+ R_\kappa (\omega_n,\vec{q})$, $M_T^R=U'(\rho)+ R_\kappa (\omega_n,\vec{q})$ and 
\begin{equation}
D =  \left[Z_\omega \omega_n^2+Z\vec{q}^2+M_L^R\right] \left[Z_\omega \omega_n^2+Z\vec{q}^2+M_T^R\right] +X^2\omega_n^2\;.
\end{equation}
Finally, $R_\kappa (\omega_n,\vec{q})$ is a regulator function added to the inverse propagator. Its particular form is specified as
\begin{equation}
R_\kappa (\omega_n,\vec{q}) = Z\left(\kappa^2-\vec{q}^2-\frac{Z_\omega}{Z}\omega_n^2\right)\theta \left(\kappa^2-\vec{q}^2-\frac{Z_\omega}{Z}\omega_n^2\right).
\end{equation} 

The quantity $U_\kappa(\rho)$ may be understood as a free energy including fluctuation modes between the momentum scale $\kappa$ and $\Lambda$. For $\kappa=\Lambda$ the fluctuations are frozen and $U_{\kappa=\Lambda}(\rho)$ is given by the MF effective potential $\omega_L(\Delta)$ [Eq.~(\ref{2})]. On the other hand, for $\kappa\to 0$ all the fluctuations are integrated and $U_{\kappa\to 0}(\rho)$ is the full free energy. Eq.~(\ref{LPA_flow}) therefore interpolates between the bare and full effective potential upon varying the momentum cutoff scale $\kappa$. It may be derived from an exact functional RG flow equation\cite{Wetterich_93} (the Wetterich equation) by neglecting renormalization of the momentum and frequency dependencies of the propagators (i.e. keeping the $Z$ and $X$ factors fixed). This constitutes the essence of the approximation. For details of the derivation see e.g. Ref.~\onlinecite{Berges_review_04}.  Observe, that Eq.~(\ref{LPA_flow}) retains the full field dependence (i.e. it does not invoke any polynomial expansion of the scale-dependent free energy $U_\kappa(\rho)$). It is therefore particularly suitable for investigating the impact of fluctuations on the order of the phase transitions. On the other hand, it presents a nonlinear partial differential equation which may be studied only numerically. We also note that simpler truncations of the Wetterich equations were applied in similar a context in Refs.~\onlinecite{Floerchinger_10, Krippa_15}.

Discretizing the $\rho$-space, we have integrated Eq.~(\ref{LPA_flow}) at $T=0$ with the initial condition given by Eq.~(\ref{2}). The obtained results indicate no signature of an instability of the QCP obtained at the MF level towards a first-order transition. We demonstrate this in Fig.~8 by plotting the MF and renormalized effective potentials for the set of parameters considered in Fig.~2,  strictly at the transition point at $T=0$ (which at MF level lies within the superfluid phase). The calculation demonstrates stability of the QCP with respect to order-parameter fluctuations in $d=3$.    
\begin{figure}[!]
\centering
    \includegraphics[width=0.45 \textwidth]{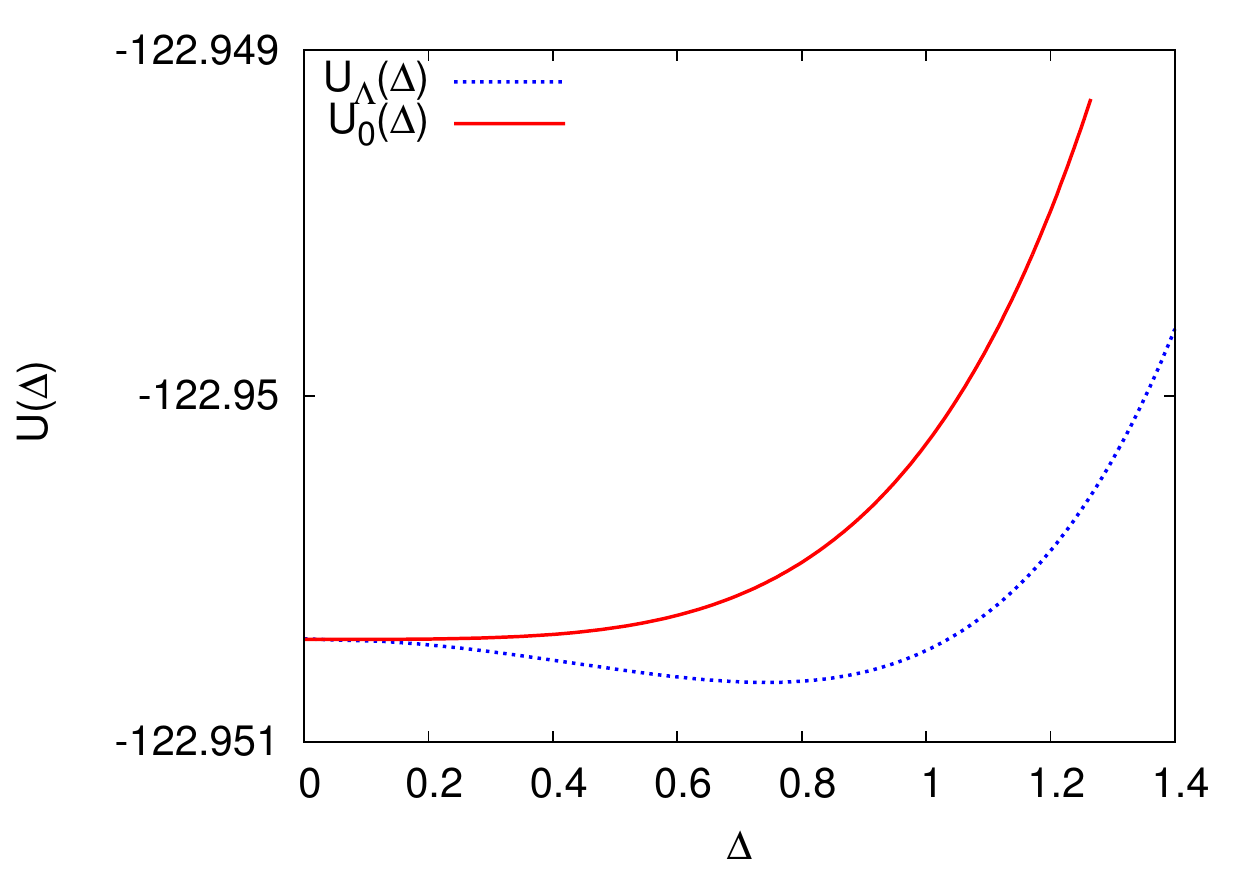}
    \caption{ The mean-field [$U_\Lambda(\Delta)$] and renormalized [$U_0(\Delta)$] effective potentials for the set of parameters considered in Fig.~2 with $T=0$ and $h= h_c\approx 1.346$. The propagator parameters are set as $Z=Z_\omega=10X=1$. The plot demonstrates a shift of the quantum critical point from $h_c^{MFT}\approx 1.362$ to $ h_c\approx 1.346$ due to order-parameter fluctuations. The transition remains second-order after including fluctuation effects via the functional RG flow. The order parameter expectation value vanishes continuously as $h$ is varied across the critical value $h_c$.}
    \label{Potentials_plot}
\end{figure}

\section{\label{sec:level5}Conclusion and outlook}
We have studied the analytical structure of the effective potential for imbalanced Fermi mixtures with particular focus on the properties of the Landau expansion at $T\to 0^+$ and the possibility of realizing a system hosting a quantum critical point. We have shown the Landau expansion to be well-defined at $T\to 0^+$ except for a subset of parameters described by Eq.~(\ref{Landau_reg}). We have demonstrated that at mean-field level the occurrence of a QCP is generally excluded in $d=2$. In $d=3$ we have found and characterized a parameter regime admitting a QCP. This is restricted to situations where one of the chemical potentials is negative so that the quantum phase transition occurs between the superfluid phase and the fully polarized gas. The second-order transition turns out to be favorable at large mass imbalance $r$. We have performed a functional RG calculation showing stability of our conclusion  with respect to fluctuation effects.  

Resolution of such quantum critical phenomena may perhaps soon appear within the range of experimental technologies bearing in mind the dynamical progress in realizing uniform Fermi gases trapped in box potentials.\cite{murkherjee_17, hueck_two-dimensional_2018} 

On the theory side, superfluid quantum criticality constitutes a largely unexplored field involving the interplay of fermionic and collective bosonic degrees of freedom. This applies to both, the uniform case considered here, as well as the hypothetical  quantum critical points in nonuniform (FFLO) superfluids.\cite{Piazza_2016, Pimenov_2017}  A complete understanding of these systems seems to pose an interesting challenge considering the interplay of a rich spectrum of  fluctuations including fermions and Goldstone modes as well as topological aspects related to the Kosterlitz-Thouless physics in $d=2$. 

\begin{acknowledgments}
We acknowledge support from the Polish National Science Center via grant 
2014/15/B/ST3/02212.
\end{acknowledgments}

\appendix

\begin{widetext}
\section{Coefficient $a_2^{(0)}$ for $d=3$}
Below we present the expressions for $a_2^{(0)}$ in $d=3$ for the six regimes illustrated in Fig.~4.

For regime A, where $\mu_\uparrow>0$ and $\mu_\downarrow>0$ we find:
\begin{eqnarray}
a^{(A)}_2&=-\frac{1}{g}-\frac{m_r}{2\pi^2}\Bigg[\Lambda +\frac{\sqrt{2\mu m_r}}{2}\ln\left(\frac{|\Lambda-\sqrt{2\mu m_r}|}{|\Lambda+\sqrt{2\mu m_r}|}\right)-\sum_\sigma\left\{\lambda_\sigma +\frac{\sqrt{2\mu m_r}}{2}\ln\left(\frac{|\lambda_\sigma-\sqrt{2\mu m_r}|}{|\lambda_\sigma+\sqrt{2\mu m_r}|}\right)\right\}\Bigg].
\label{A1}
\end{eqnarray}
For the regions B and C  (where $\mu>0$, $\mu_\sigma>0$ and $\mu_{\bar{\sigma}}<0$) we obtain:
\begin{eqnarray}
a^{(B,C)}_2&=-\frac{1}{g}-\frac{m_r}{2\pi^2}\Bigg[\Lambda +\frac{\sqrt{2\mu m_r}}{2}\ln\left(\frac{|\Lambda-\sqrt{2\mu m_r}|}{|\Lambda+\sqrt{2\mu m_r}|}\right)-\lambda_\sigma -\frac{\sqrt{2\mu m_r}}{2}\ln\left(\frac{|\lambda_\sigma-\sqrt{2\mu m_r}|}{|\lambda_\sigma+\sqrt{2\mu m_r}|}\right)\Bigg].
\label{A2}
\end{eqnarray}
For the regions D and E ($\mu<0$, $\mu_ \sigma>0, \mu_{\bar{\sigma}}<0$)  we have:
\begin{eqnarray}
a^{(D,E)}_2&=-\frac{1}{g}-\frac{m_r}{2\pi^2}\Big[\Lambda -\sqrt{2\bar{\mu} m_r}\arctan\left(\frac{\Lambda}{\sqrt{2\bar{\mu} m_r}}\right)-\lambda_\sigma +\sqrt{2\bar{\mu} m_r}\arctan\left(\frac{\lambda_\sigma}{\sqrt{2\bar{\mu} m_r}}\right)\Big]\;,
\label{A3}
\end{eqnarray}
where $\bar{\mu}=-\mu$. 

Finally, for the subset F ($\mu_ \sigma,\mu_{\bar{\sigma}}< 0$) we obtain the following expression:
\begin{eqnarray}
a^{(F)}_2=-\frac{1}{g}-\frac{m_r}{2\pi^2}\Bigg[\Lambda -\sqrt{2\bar{\mu} m_r}\arctan\left(\frac{\Lambda}{\sqrt{2\bar{\mu} m_r}}\right)\Bigg].
\label{A4}
\end{eqnarray}
\section{Coefficient $a_4^{(0)}$ for $d=3$}
Here we present the expressions for $a_4^{(0)}$ in $d=3$ for the six regimes illustrated in Fig.~4.

For the region A ($\mu_\uparrow>0$ and $\mu_\downarrow>0$) we obtain:
\begin{eqnarray}
a_4^{(A)}=-\frac{m_r^2}{32\pi^2}\Bigg[\frac{\Lambda\left(\Lambda^2+2\mu m_r\right)}{\mu\left(\Lambda^2-2\mu m_r\right)^2}+\frac{m_r}{(2\mu m_r)^{3/2}}\ln\left(\frac{|\Lambda-\sqrt{2\mu m_r}|}{|\Lambda+\sqrt{2\mu m_r}|}\right)+\nonumber\\
-\sum_\sigma\left(\frac{
\lambda_\sigma\left(\lambda_\sigma^2+2\mu m_r\right)}{\mu\left(\lambda_\sigma^2-2\mu m_r\right)^2}+\frac{m_r}{(2\mu m_r)^{3/2}}\ln\left(\frac{|\lambda_\sigma-\sqrt{2\mu m_r}|}{|\lambda_\sigma+\sqrt{2\mu m_r}|}\right)-\frac{8m_\sigma\lambda_\sigma}{\left(\lambda_\sigma^2-2\mu m_r\right)^2}\right)\Bigg]\;.
\label{B1}
\end{eqnarray}
For the regions B and C (where $\mu>0$, $\mu_\sigma>0$ and $\mu_{\bar{\sigma}}<0$)  the expression for $a_4^{(0)}$ is given by:
\begin{eqnarray}
a_4^{(B-C)}=-\frac{m_r^2}{32\pi^2}\Bigg[\frac{\Lambda\left(\Lambda^2+2\mu m_r\right)}{\mu\left(\Lambda^2-2\mu m_r\right)^2}+\frac{m_r}{(2\mu m_r)^{3/2}}\ln\left(\frac{|\Lambda-\sqrt{2\mu m_r}|}{|\Lambda+\sqrt{2\mu m_r}|}\right)+\nonumber\\
-\frac{\lambda_\sigma\left(\lambda_\sigma^2+2\mu m_r\right)}{\mu\left(\lambda_\sigma^2-2\mu m_r\right)^2}-\frac{m_r}{(2\mu m_r)^{3/2}}\ln\left(\frac{|\lambda_\sigma-\sqrt{2\mu m_r}|}{|\lambda_\sigma+\sqrt{2\mu m_r}|}\right)+\frac{8m_\sigma\lambda_\sigma}{\left(\lambda_\sigma^2-2\mu m_r\right)^2}\Bigg]\;.
\label{B2}
\end{eqnarray}
For the regions D and E ($\mu<0$, $\mu_ \sigma>0, \mu_{\bar{\sigma}}<0$) we have:
\begin{eqnarray}
a_4^{(D-E)}=\frac{m_r^2}{32\pi^2}\Bigg[\frac{\Lambda\left(\Lambda^2-2\bar{\mu} m_r\right)}{\bar{\mu}\left(\Lambda^2+2\bar{\mu} m_r\right)^2}+\frac{2m_r}{(2\bar{\mu} m_r)^{3/2}}\arctan\left(\frac{\Lambda}{\sqrt{2\bar{\mu} m_r}}\right)+\nonumber\\
-\frac{\lambda_\sigma\left(\lambda_\sigma^2-2\bar{\mu} m_r\right)}{\mu\left(\lambda_\sigma^2
+2\bar{\mu} m_r\right)^2}-\frac{2m_r}{(2\bar{\mu} m_r)^{3/2}}\arctan\left(\frac{\lambda_\sigma}{\sqrt{2\bar{\mu} m_r}}\right)-\frac{8m_\sigma\lambda_\sigma}{\left(\lambda_\sigma^2+2\bar{\mu} m_r\right)^2}\Bigg].
\label{B3}
\end{eqnarray}
Finally, for region F the Landau coefficient $a_4^{(0)}$ is given by:
\begin{eqnarray}
a_4^{(F)}=\frac{m_r^2}{32\pi^2}\Bigg[\frac{\Lambda\left(\Lambda^2-2\bar{\mu} m_r\right)}{\bar{\mu}\left(\Lambda^2+2\bar{\mu} m_r\right)^2}
+\frac{2m_r}{(2\bar{\mu} m_r)^{3/2}}\arctan\left(\frac{\Lambda}{\sqrt{2\bar{\mu} m_r}}\right)\Bigg].
\label{B4}
\end{eqnarray}

\end{widetext}

%

\end{document}